
\magnification 1200
\baselineskip=14pt
\null\vskip -0.5truein
\hsize=6.0truein
\vsize=8.5truein
\font\tenrm=cmr8
\centerline{\bf SEARCHING FOR SUSY DARK MATTER}
\bigskip
\centerline{R. ARNOWITT\footnote\dag{Conference Speaker}}
\centerline{Center for Theoretical Physics, Department of Physics}
\centerline{Texas A\&M University, College Station, Texas  77843-4242}
\smallskip
\centerline{and}
\smallskip
\centerline{PRAN NATH}
\centerline{Department of Physics, Northeastern University}
\centerline{Boston, MA  02115}
\bigskip
\centerline{ABSTRACT}
\medskip
\leftskip=3pc
\rightskip=3pc
\baselineskip=12pt
{\tenrm The possibility of detecting supersymmetric dark matter is
examined within the framework of the minimal supergravity model (MSGM),
where the $\tilde{Z}_{1}$ is the LSP for almost the entire parameter
space.  A brief discussion is given of experimental strategies for
detecting dark matter.  The relic density is constrained to obey 0.10
$\leq \Omega_{\tilde{Z}_{1}}h^2 \leq$0.35, consistent with COBE data.
Expected event rates for an array of possible terrestial detectors
($^3$He, CaF$_2$, Ge, GaAs, NaI and Pb) are examined.  In general,
detectors relying on coherrent $\tilde{Z}_{1}$-nucleus scattering are
more sensitive than detectors relying on incoherrent (spin-dependent)
scattering.  The dependence of the event rates as a function of the
SUSY parameters are described.  The detectors are generally most
sensitive to the small $m_0$ and small $m_{\tilde{q}}$ and large
tan$\beta$ part of the parameter space.  The current $b\rightarrow s+\gamma$
decay rate eliminates regions of large event rates for $\mu >0$,
but allows large event rates to still occur for $\mu<0$.  MSGM
models that also possess SU(5)-type proton decay generally predict
event rates below the expected sensitivity of current dark matter
detectors.}
\bigskip
\leftskip=0pc
\rightskip=0pc
\baselineskip=14pt
\noindent
{\bf 1.  Introduction}
\vskip 0.4cm
If the SUSY models currently being examined are correct, then
supersymmetry will be discovered at the LHC in the year 2005 or
possibly even at an upgraded version of the Tevatron (e.g. the
DiTevatron) in the year 2000.  However, high energy colliders may not
shed further light until then.  Thus it is of interest to look at other
phenomena which supersymmetry might effect, e.g. dark matter, proton
decay, the $b\rightarrow s+\gamma$ decay etc.  Each of these restricts
the parameter space of supersymmetry, and so by combining the
constraints, one can get sharper predictions of what to expect at
colliders.  We will discuss here the question of detecting SUSY dark
matter, and how bounds on other processes affect dark matter searches.
There is a warning however one should make concerning such analyses:
In applying SUSY to dark matter searches, one is making {\it
additional} assumptions (e.g., cosmological assumptions) not made in
dealing with accelerator phenomena.  We will test the sensitivity of
final results to some of these extra assumptions, but some caution is
needed in interpreting the theoretical predictions.
\vskip 0.6cm
\noindent
{\bf 2.  Dark Matter}
\vskip 0.4cm
There is much astronomical evidence that more than 90\% of the total
mass of the Galaxy and perhaps of the universe is made up of dark
matter of an unknown type.  In the Galaxy this can be seen from
rotation curves of luminous matter (Fig. 1).  The circular velocity
$v_{cir}$ does not fall with $r$ beyond the optical radius.  Similar
\vskip230pt
\noindent
Fig. 1.  Estimated mean spherical density of dark matter in the Galaxy$^1$.
\vskip 0.1cm
\noindent
effects are seen in other galaxies (Fig. 2).  In the vicinity of the
sun, the mean density of dark matter (DM) is estimated as

$$\rho_{DM}\cong 0.3~~GeV/cm^{3}\eqno(1)$$
\noindent
 (about $10^{4}\gamma_{universe}$) and assuming a Maxwell velocity
 distribution, the DM has a velocity relative to the solar system of

$$v_{DM}\cong 320   km/s\eqno(2)$$

\vfill\eject
\noindent
i.e. $v_{DM}/c\approx 10^{-3}$.

\vskip195pt
\noindent
Fig. 2  $v_{cir}$ for a number of galaxies showing $v_{cir}$ remains
approximately constant well beyond the optical radius$^2$.
\vskip 0.1cm
There  are a large number of candidates for dark matter, both from
astronomy and particle physics.  Supersymmetric models with R-parity
offer two candidates:  the lightest neutralino $\tilde {Z}_{1}$ and the
sneutrino, $\tilde{\nu}$.  However, in supergravity models almost
always the $\tilde{Z}_{1}$ is the lightest supersymmetric particle
(LSP) and hence is absolutely stable.  Thus the relic $\tilde{Z}_{1}$,
left over from the big bang would be the dark matter.  The dynamics is
then well fixed and we will deal with this case exclusively here.

COBE data suggests that DM is a mix of cold dark matter (CDM) [which we
are assuming here to be the $\tilde{Z}_{1}$] and hot dark matter (HDM)
[possibly massive neutrinos] in the ratio of about 2:1.  There can also
be baryonic dark matter (B) but nucleosynthesis analyses limit this to
$\buildrel{<}\over{\sim}$ 10\%.  Thus if we define
$\Omega_{i}=\rho_{i}/\rho_{c}$, where $\rho_{i}$ is the mass density of
type $i$ and $\rho_{c}=3H^{2}/(8\pi G_{N})$ is the critical mass
density to close the universe (H=Hubble constant and $G_{N}$=Newtonian
gravitational constant), the inflationary scenario requires
$\Sigma\Omega_{i}=1$ and hence a reasonable mix of matter is
$$\Omega_{\tilde{Z}_{1}}\simeq 0.6; ~~ \Omega_{HDM}\simeq 0.3; ~~
  \Omega_{B}\simeq 0.1\eqno(3)$$

What can be calculated theoretically is $\Omega_{\tilde{Z}_{1}}h^{2}$
where h=H/(100 km/s Mpc).  Astronomical observations give a range of
values for h i.e. $h\cong 0.5-0.75$.  Hence

$$\Omega_{\tilde {Z}_{1}}h^{2}\cong 0.10-0.35\eqno(4)$$
\noindent
This bound strongly restricts the SUSY parameter space.  We will
discuss below how sensitive our results are to the endpoints of this
bound on $\Omega_{\tilde{Z}_{1}}h^{2}$.
\vskip 0.6cm
\noindent
{\bf 3.  Detection Strategies}
\vskip 0.4cm
The solar system is presumably being bombarded with $\tilde{Z}_{1}$
particles moving with velocity $<v_{\tilde{Z}_{1}}>\cong$ 320 km/s.
Two strategies have been proposed for their detection.
\vskip 0.6cm
\noindent
{\it 3.1 Indirect Detection}
\vskip 0.4cm
The $\tilde{Z}_{1}$ impinging on the sun can be relatively easily
captured since the escape velocity at the surface of the sun is 618
km/s$\cong \nu_{\tilde{Z}_{1}}$.  Once captured the $\tilde{Z}_{1}$
will be slowed down inside the sun and gradually fall to the center.
There they accumulate and can annihilate giving rise to neutrinos via e.g.
${\tilde{Z}_{1}}+{\tilde{Z}_{1}}
\rightarrow b+\bar b,....\rightarrow\nu_{\mu}+X$.
Since $m_{\tilde{Z}_{1}}\cong$(10-100) GeV, high energy neutrinos
coming from the sun from this process would be a striking signal that
could be observed on Earth by a neutrino telescope.  Calculations
indicate that one would need a telescope of area $> 1km^2$ to cover the
SUSY parameter space$^3$, and telescopes of this size are currently
being built.

\vskip 0.6cm
\noindent
{\it 3.2 Direct Detection}
\vskip 0.4cm
A direct approach to see the incoming $\tilde{Z}_{1}$ is to detect
their scattering by quarks in nuclei of a terrestrial detector:
${\tilde{Z}_{1}}+q\rightarrow \tilde{Z}_{1}+q$.  Two types of detectors
being considered are the following.
\vskip 0.6cm
\noindent
{\it 3.21  Low Temperature Detectors}
\vskip 0.4cm
Since $m_{\tilde{Z}_{1}}\simeq$ (10-100) GeV, the recoil energy to a
nucleus that has been struck is $\Delta E\approx
(v_{\tilde{Z}_{1}}/c)^2(m_{\tilde{Z}_{1}}c^2)\simeq$ (10-100) KeV.
This is of a size to produce phonons (heat) or ionization in the
lattice of the detector.  The temperature rise $\Delta$T is
$\Delta$T=$\Delta$E/{VC} where C is the specific heat and V the
volume.  Since at low temperature C$\sim$T$^{3}$ one can enhance
$\Delta$T by reducing the temperature but one is also limited by V not
being too large.  An optimum set of parameters is to have T in the mK
range and a detector of mass of $\sim$1kg.  Detectors of this type are
currently being built.
\vskip 0.6cm
\noindent
{\it 3.22  Superconducting Detectors}
\vskip 0.4cm
Here superconducting granules are suspended in a dielectric carrier in
the presence of a magnetic field (Fig. 3.)  The superconductors are put
into a metastable
\vskip135pt
\noindent
Fig. 3  Schematic diagram of a superconducting detector$^1$.
\vskip 0.1cm
\noindent
state and hence the magnetic field is excluded from the granules by the
Meissner effect.  When a $\tilde{Z}_{1}$ strikes a granule, the
deposited energy triggers the transition to the normal state, and the
magnetic flux movement then produces a signal in the pickup coil.  The
characteristic size of such detectors are also about 1 kg.
\vskip 0.1cm
Background for these DM detectors include cosmic ray muons and natural
radioactivity.  The present sensitivity expected is 0.1 events/kg da,
and this might be improved at a later date (i.e. by going underground)
to 0.01 events/kg da.
\vskip 0.6cm
\noindent
{\bf 4.  Dynamical Model}
\vskip 0.4cm
To calculate the event rates expected at terrestial detectors, we need
to calculate two items:  (1) the relic density of the $\tilde{Z}_{1}$,
in order to make sure that $\Omega_{\tilde{Z}_{1}}h^{2}$ lies in the
range of Eq. (4), and (2) the $\tilde{Z}_{1}$-nucleus cross section to
obtain the expected event rate for a given detector.  The bounds on
$\Omega_{\tilde{Z}_{1}}h^2$ significantly limits the SUSY parameter
space and hence strongly affects the event rates obtained from the
$\tilde{Z}_{1}$-nucleus cross sections.

In order to carry out the above calculations one needs a dynamical
model and we use here models based on supergravity grand
unification$^4$.  These models have the advantages of (i) being
consistent with the LEP data on unification of couplings at $M_G\simeq
10^{16}$ GeV, (ii) generating spontaneous breaking of supersymmetry at
M$_G$ (in the ``hidden'' sector), (iii) generating naturally
spontaneous breaking of SU(2)$\times$U(1) at the electroweak scale by
radiative corrections, and (iv) having all new SUSY phenomena described
by only 4 new extra parameters and one sign.

The minimal supergravity model (MSGM) is characterized at the Gut scale
by a superpotential

$$W=\mu_{0}H_{1}H_{2}+W_{Y}+{1\over{M_{G}}}W^{(4)}\eqno(5)$$
\noindent
where W$_Y$ is the cubic Yukawa couplings and W$^{(4)}$ is the quartic
non-renormalizable couplings (possibly leading to proton decay).  In
addition there is a soft supersymmetry breaking effective potential
$$V_{SB}=m_{0}^{2}{\sum_{a}}{z_{a}}^{+}z_{a}
  +(A_{0}W_{Y}+B_{0}\mu_{0}H_{1}H_{2}+h.c.)\eqno(6)$$
\noindent
where $\{z_a\}$ are the scalar fields, and a universal gaugino mass
term $L_{mass}^{\lambda}=-m_{1/2}\bar\lambda^{\alpha}\lambda^{\alpha}$
where $\lambda^{\alpha}$ are the gaugino fields.  The scalar mass $m_0$
(and cubic soft breaking constant $A_0$) are universal provided the
agent of supersymmetry breaking in the hidden sector (e.g. the super
Higgs field) communicates with the physical sector in a flavor
independent way.  This is automatically the case for contributions
arising from the effective potential (since there the only
communication is gravitational) and will be true in general if the
couplings of the superHiggs to the physical fields in the Kahler
potential is also flavor independent.  Eq. (6) guarantees a natural
suppresion of unwanted FCNI.

Using the renormalization group equations (RGE) one obtains at the
electroweak scale the conditions for SU(2)$\times$U(1) breaking

$${1\over 2} M_{Z}^{2}=-\mu^{2}+{{{m_{{H_1}}^2}{-m_{H{_2}}^2 tan^2\beta}}
  \over{tan^2\beta-1}};~~
sin2\beta = -{{{B\mu}\over{{2\mu^2}+{m_{H_1}^{2}}+{m^{2}_{H}}{_2}}}}
\eqno(7)$$

\noindent
where $tan\beta\equiv<H_{2}>/<H_{1}>$ and $m_{H{_i}}^2$ are the Higgs
running masses including 1-loop corrections.  One can then eliminate
$B_{0}$ and $\mu_{0}^{2}$ leaving four parameters $m_{0}$, $m_{1/2}$,
$A_{t}$ (the t-quark A parameter at the electroweak scale), tan$\beta$
and the sign of $\mu_{0}$ to determine all the masses and interactions
of the 32 new SUSY particles.  One therefore expects a large number of
relations holding between the SUSY masses.  If one limits the parameter
space so that (i) experimental mass bounds of LEP and the Tevatron are
obeyed, (ii) $m_{0}$, $m_{\tilde{g}}<$ 1 TeV, $m_{\tilde{g}}$ the
gluino mass (so that no extreme fine tuning of parameters occurs), and
(iii) radiative breaking of SU(2)$\times$U(1) occurs at the electroweak
scale, then the following ``scaling'' relations hold throughout most of
the parameter space$^5$:  2m$_{\tilde{Z}_{1}}\cong m_{\tilde{Z}_{2}}\cong
m_{\tilde{W}_{1}}\simeq({1\over3}-{1\over4})m_{\tilde{g}}$;
$m_{\tilde{Z}_{3}}\cong m_{\tilde{Z}_{4}}\cong
m_{\tilde{W}_{2}}>>m_{\tilde{Z}_{1}}$; and $m_{H{^0}}\cong
m_{{A}^{0}}\cong m_{{H}^{\pm}}>>m_{h}.$ Here the $\tilde{W}_{i}$ are
the charginos, the $\tilde{Z}_{i}$ are the neutralinos, $h^0$ and $H^0$
are the CP even Higgs, $A^0$ is the CP odd Higgs and $H^\pm$ is the
charged Higgs.  In addition one always has tan$\beta>1$.
\vskip 0.6cm
\noindent
{\bf 5.  Calculation of $\tilde{Z}_{1}$ Relic Density}
\vskip 0.4cm
As discussed above, Eq. (4) puts a significant constraint on the
allowed region of parameter space.  Since this band is relatively
narrow, it is important to include major effects in calculating the
relic density.

R parity makes the $\tilde{Z}_{1}$ produced in the early universe
absolutely stable.  How-
\vskip140pt
\noindent
Fig. 4  Annihilation diagrams of the $\tilde{Z}_{1}$ in the early universe.
\vskip 0.1cm
\noindent
ever, they can annihilate in pairs and the main annihilation diagrams
are shown in Fig. 4.  The calculation of the mass density of the
$\tilde{Z}_{1}$ remaining at the present time proceeds as follows$^6$:
Initially the $\tilde{Z}_{1}$ are in thermal equilibrium with the
background and the reactions of Fig. 4 go forward and backward.
However, when the annihilation rate falls below the expansion rate of
the universe, ``freezeout'' occurs at temperature $T_{f}$, and the
$\tilde{Z}_{1}$ disconnect from the background.  The $\tilde{Z}_{1}$
then continue to annihilate and the amount left at present time is$^6$

$$\Omega_{\tilde{Z}_{1}}h^{2}\cong 2.4\times 10^{-11}
{\left({{{T_{\tilde{Z}_{1}}}\over{T_{\gamma}}}}\right )}^{3}
{\left({{T_{\gamma}}\over 2.73}\right)}^{3}
{{N_{f}}\over{J(x_{f})}}\eqno(8)$$
\noindent
where $N_f$ is the effective number of degrees of freedom,

$$J(x_f)=\int_{0}^{x_f}<\sigma v>dx; ~~  x=kT/m_{\tilde{Z}_{1}},\eqno(9)$$
\noindent
and $<\sigma v>$ is the thermal average ($\sigma$ = annihilation cross
section, $v$ = relative velocity):

$$<\sigma v>=\int_{0}^{\infty}dvv^{2}(\sigma v)Exp(-v^{2}/4x)
  \int_{0}^{\infty}dvv^{2}Exp(-v^{2}/4x)\eqno(10)$$
\noindent
In general, the annihilation occurs non-relativistically, i.e.
$x_f\approx {1\over{20}}$.  However, this does not mean one can always
make a non-relativistic expansion, $\sigma v=a+bv^{2}+...$, in
performing the calculation of J(x$_f$).  As has been pointed out$^7$,
such an expansion fails near an s-channel pole.  For example for the h
pole one has

$$\sigma v=A{{(s-4 m_{\tilde{Z}_{1}}^2)/m_{\tilde{Z}_{1}}^2}
  \over{(s-m_{h}^{2})^{2}+m_{h}^{2}\Gamma_{h}^{2}}}\eqno(11)$$
\noindent
where $\Gamma_{h}$=O(MeV) is the h boson width and A is a constant.
Since $\Gamma_h$ is small one must treat the pole more carefully.  The
danger of not doing so is shown in Fig. 5$^8$.
\vskip150pt
\noindent
Fig. 5.  $\Omega_{approx}$ is the $\tilde{Z}_{1}$ relic density
calculated in the $\sigma v=a+bv^2$ approximation and $\Omega$ is the
relic density calculated rigorously using Eq. (11).  The h and Z poles
occur at the points where the curve descends through zero.
\vskip 0.1cm
\noindent
Note the long tail where an error of a factor of $\approx$ 2 can be
made well past the h and Z poles.  One finds, for example, that
$\Omega_{approx}$ has an error of $>$ 25\% for 65\% of the parameter
points where $m_{\tilde{g}}<450$ GeV, while $\Omega_{approx}$ is a good
approximation for $\simeq$ 100\% of the points for
$m_{\tilde{g}}\geq 450$ GeV.  The reason for this is the scaling
relations discussed at the end of Sec. 4.  Since
$2m_{\tilde{Z}_{1}}>({1\over 3}-{1\over 4})m_{\tilde{g}}$, when
$m_{\tilde{g}}\geq 450$ GeV, the $\tilde{Z}_{1}$ has passed both the h
and Z poles (where $2m_{\tilde{Z}_{1}}\simeq m_{h}$ or $M_Z$).  (Recall
that in the  MSGM one has $m_{h}< 130$ GeV.)  However, for lighter
$m_{\tilde{g}}$ the pole effects are very important, since one is
almost always near an h or a Z pole.
\vskip 0.6cm
\noindent
{\bf 6.  Detector Rates}
\vskip 0.4cm
The dark matter detectors discussed in Sec. 3.2 detect the
$\tilde{Z}_{1}$ from their scattering by quarks in the nucleus.  The
basic diagrams are given in Fig. 6.  They
\vskip130pt
\noindent
Fig. 6  $\tilde{Z}_{1}$-quark scattering diagrams.  $\tilde{q}$ is the squark.
\vskip 0.1cm
\noindent
are mainly crossed diagrams to the relic annihilation diagrams of Fig.
4.  Thus to a rough approximation, when the relic density of the
$\tilde{Z}_{1}$ is small (i.e. there is a large annihilation cross
section) the number of scattering events will be large.  This makes the
results somewhat sensitive to the lower bound,
$\Omega_{\tilde{Z}_{1}}h^2\geq 0.1$, that we have chosen, and we will
discuss this sensitivity below in Sec. 8.

Calculations show that it is possible to represent the $\tilde{Z}_{1}$-q
scattering by an effective Lagrangian$^9$:

$$L_{eff}=(\bar \chi_{1}\gamma^{\mu}\gamma^{5}\chi_{1})
  \bar q\gamma^{\mu}(A_{q}P_{L}+B_{q}P_{R})q
+(\bar \chi_{1}\chi_{1})C_{q}m_{q}\bar q q\eqno(12)$$
\noindent
where $\chi_{1}$(x) is the $\tilde{Z}_{1}$ field, q(x) the quark field,
and $m_q$ its mass and $P_{R,L}={1 \over 2}(1\pm\gamma^5)$.  The
coefficients $A_q$, $B_q$ arise from the Z t-channel and $\tilde{q}$
s-channel poles while $C_q$ arises from the h and H t-channel and
$\tilde{q}$ s-channel poles.

The first term of Eq. (12) leads to spin dependent incoherrent
scattering by the quarks in the nucleus, while the second term leads to
coherrent scattering, where the masses of all the quarks (and hence
nucleons) approximately add coherrently.  Thus the coherrent amplitude
will contain a factor $M_N$ larger then the incoherrent amplitude,
where $M_N$ is the nucleus mass.

In general, the $\tilde{Z}_{1}$ field is a mix of higgsinos and gauginos:

$$\chi_{1}=n_{1}\tilde{W}_{3}+n_{2}\tilde{B}+n_{3}\tilde{H}_{2}
  +n_{4}\tilde{H}_{1}\eqno(13)$$
\noindent
where $\tilde{W}_{3}$ and $\tilde{B}$ are the $W_3$ and U(1) [Bino]
gauginos, and $\tilde{H}_{1,2}$ are Higgsinos.  The $n_{i}$ can be
computed in terms of the basic SUSY parameters $m_0$, $m_{1 \over 2}$,
$A_t$ and tan$\beta$.  Over most of the allowed part of the parameter
space one finds

$$n_{2}>n_{1},n_{4}>>n_{3}\eqno(14)$$
\noindent
In SUSY theory, $m_h$ is small i.e. $m_{h}<(120-130)$ GeV, and
$m_{H^{2}}>>m_{h^{2}}$.  In spite of this it is remarkable that the H
contribution is important in the coherrent amplitude$^{10}$.  The
reason for this is the following.  One finds for $C_q$ the result

$${C_q} = {{{g_2^2}\over{4M_W{^2}}}\left[\left\{{{{{{cos\alpha}\over
{sin\beta}}\atop-{{sin\alpha}\over
{cos\beta}}}}{{{F_h}\over{m_h^2}}\atop{{F_h}\over{m_h^2}}}}\right\}
+\left\{{{{{sin\alpha}\over{sin\beta}}\atop{{cos\alpha}\over{cos\beta}}}}
{{{F_H}\over{m_H^2}}\atop{{F_H}\over{m_H^2}}}\right\}\right]
{u-quark\atop d-quark}} \eqno(15)$$

\noindent where $\alpha$ is the rotation angle needed to diagonalize
the 2$\times$2 h-H mass matrix,
$F_{h}=(n_{1}-n_{2}tan\theta_{W})(n_{3}cos\alpha-n_{4}sin\alpha )$ and
$F_{H}=(n_{1}-n_{2}tan\theta_{W})(n_{3}sin\alpha-n_{4}cos\alpha )$.
In calculating the h-H mass matrix, one must include as is well-know,
the one loop corrections due to the fact that $m_{t}$ is large.  One
finds then, for the allowed part of the parameter space that $\alpha$
is  generally small i.e. $\alpha\approx{1\over10}$.  This result
combined with Eq. (14) shows that generally cos$\alpha F_{H}>>sin\alpha
F_{h}$, which can overcome the reduction in the d-quark amplitude due
to the largeness of $m_{H}$.  One finds, in general, as one varies over
the parameter space that the H contribution to $C_q$ can vary from
${1\over{10}}$ to 10 times the h contribution for d-quarks, but is
generally a small correction for u-quarks.
\vskip 0.6cm \noindent {\bf 7.  Detector Event Rates}
\vskip 0.4cm
The total event rate expected for a given dark matter detector can be
written in the following form$^9$

$$R=[R_{coh}+R_{inc}][{{\rho_{\tilde {Z}_1}\over 0.3 GeV cm^{-3}}]
[{\langle v_{\tilde {Z}_1}\rangle \over 320km/s}]
{events\over kg~da}}\eqno(16)$$
\noindent
where

$$R_{coh}={4m_{\tilde {Z}_1}M{_N^3}M{_Z^4}\over[M_N+m_{\tilde {Z}_1}]^2}
  |A_{coh}|^2\eqno(17)$$

$$R_{inc}={4m_{{\tilde {Z}_1}M_N}\over{[M_N+m_{\tilde {Z}_1}]^2}}\lambda^2
  J(J+1)|A_{inc}|^2\eqno(17a)$$
\noindent
Here ~~$A_{coh}\sim C_{q},~~A_{inc}\sim B_{q}-A_{q}$,~~J is the nuclear spin
and~ ~$\lambda~<~N|{\buildrel\rightarrow\over{J}}|N~>$
{}~~=~~~$<N|\Sigma{\buildrel\rightarrow\over{S}}_{i}|N>$
where the sum is over the spins of all nucleons in nucleus N.  Note
that for large $M_N$, $R_{coh}$ increases as $M_{N}$ while $R_{inc}$
decreases as 1/$M_{N}$.  This additional $M_{N}^{2}$ factor in
$R_{coh}$ is as expected from the discussion following Eq. (12).

We have examined the expected detector event rates for detectors made
from the following nuclei$^{11}$:~~~
$^3$He, $^{40}$Ca $^{19}F_{2}$, ~~$^{76}$Ge + $^{73}$Ge~(equal mix of
isotopes), ~~ $^{73}$Ga $^{75}$As, $^{23}$Na $^{127}$I,
and $^{207}$Pb.  Both $^{19}$F and $^3$He have strong spin
interactions, while Ge, I and Pb are increasingly heavy and hence have
increasingly strong coherrent scattering.

The parameter space studied was

$$100 GeV \leq {m_{0}}, {m_{\tilde{g}}} \leq 1 TeV; -6 \leq A_{t}/m_{0}\leq 6;
  2 \leq {tan\beta} \leq {20}\eqno(18)$$
\noindent
and the mesh used was $\Delta m_{0}$=100 GeV, $\Delta m_{\tilde{g}}$=25
GeV, $\Delta A_{t}/m_{0}=0.5$, $\Delta (tan\beta)$=2, 4, and the t
quark mass was set at $m_{t}$=167 GeV.

The dependence of the event rates on the SUSY parameters is generally
quite complicated.  However, it is possible to understand these
dependences in a qualitative way.  Fig. 7 shows that event rates
decrease rapidly with $m_{\tilde{g}}$.  This follows from the fact that
the $\tilde{Z}_{1}$ becomes increasingly Bino as $m_{\tilde{g}}$
increases i.e. $n_2$ of Eq. (13) grows and hence $n_3$ and $n_4$
shrinks, making the interference between the gaugino and Higgsino in
$F_h$ and $F_H$ (needed for coherrent scattering) become small.
$CaF_{2}$ has the strongest spin dependent forces while Pb is the
heaviest of the detectors
chosen.  One sees from this that $R_{coh}$
is significantly larger than $R_{inc}$, a general feature.

Fig. 7 also shows that R increases with tan$\beta$, a feature which can
be seen in detail in Fig. 8.  This behavior follows from the
1/$\cos\beta\sim$tan$\beta$ factor for d-quark scattering in Eq. (15).
One sees again the scaling of R with $M_{N}$ i.e. the
$^{23}$Na$^{127}$I curves lie above the $^{73}$Ge+$^{76}$Ge curves.
The dependence of R on $m_0$ is somewhat complicated.  One expects R to
decrease with $m_0$ since $m_{\tilde{q}}^2$ increases with $m_{0}^{2}$
and hence the squark s-channel contribution of Fig. 6.  shrinks.  In
addition, however, the $\mu^{2}$ determined by the radiative breaking
condition Eq. (7) increases as $m_{0}^{2}$ increases, making the
$\tilde{Z}_{1}$ increasingly more Bino like which further reduces
$R_{coh}$.
\vfill\eject
\noindent
This decrease in R with increasing $m_0$ shown in Fig. 9.
\vskip165pt
\noindent
Fig. 7  Event rate as a function of $m_{\tilde{g}}$ for Pb (solid) and
Ca $F_{2}$ (dashed) detectors.  The upperline in each pair is for
tan$\beta$=20 and lower line for tan$\beta$=6.  The curves are for
$A_{t}/m_{0}$=1.5, $m_{0}$=100 GeV, $\mu>0$.

\vskip170pt
\noindent
Fig. 8  Event rates for NaI and Ge detectors vs. tan$\beta$ for
$m_{\tilde{g}}$=275 GeV.  The dot dash curve is for $A_{t}/m_{0}$=1.0,
$m_{0}$=200 GeV, the dashed curve for $A_{t}/m_{0}$=0.5, $m_{0}$=300
GeV, and the solid curve for $A_{t}/m_{0}$=0.0, $m_{0}$=200 Gev,
$\mu>0$.  The upper curve of each pair is for NaI, the lower for Ge.
The $A_{t}, m_{0}$ parameters were chosen so that
$\Omega_{\tilde{Z}_{1}}h^2$ is approximately the same in each case of a
fixed tan$\beta$.

\vfill\eject
\noindent
The $\tilde{Z}_{1}$ mass is an increasing function of $m_{\tilde{g}}$
and as $m_{\tilde{g}}$ increases, the relic density
\vskip160pt
\noindent
Fig. 9.  Event rate as a function of $m_0$ for $m_{\tilde{g}}$=300 GeV,
$A_{t}/m_{0}$=0.5, tan$\beta$=8, $\mu>0$.  The solid curves from bottom
to top are for Ge, NaI and Pb and the dashed curve is for CaF$_2$.
\vskip 0.1cm
\noindent
increases (i.e. the annihilation cross section in the early universe
decreases).  The upper bound on $\Omega_{\tilde{Z}_{1}}h^2$ then leads
to an upper bound on $m_{\tilde{g}}$.  In general this upper bound on
$m_{\tilde{g}}$ of about 750 GeV.  This effect is shown in Fig. 10.
\vskip180pt
\noindent
Fig. 10  $(m_{\tilde{g}}$)$_{max}$ vs. tan$\beta$, $\mu>0$.  The
curves, from bottom to top are for ~~$A_{t}/m_{0}$~~~=~~ -0.5, 0.5,
1.0, 2.0, 3.0 and 3.5.
\vskip 0.2cm
One may scan the entire parameter space to obtain the maximum and
minimum event rates as a function of $A_{t}$.  These are shown in Fig.
11 for $\mu>0$ and Fig. 12 for $\mu<0$ for the domain $2\leq
tan\beta\leq 20$.  As expected from Fig. 8, the maximum event rates
occur for tan$\beta$=20.  However, the minimum rates occur at different
tan$\beta$ for different $A_t$.  Current dark matter detectors can
achieve a sensitivity of R$\simeq$0.1
\vskip180pt
\noindent
Fig. 11  Maximum and minimum event rates as a function of $A_{t}$ for
$\mu<0$ for Pb (solid) and CaF$_2$ (dashed) detectors.
\vskip180pt
\noindent
Fig. 12  Same as Fig. 11 for $\mu>0$.
\vskip 0.1cm
\noindent
 event/kg da with perhaps a factor of 10 improvement in future
sensitivity.  One sees that only part of the parameter space i.e. the
region with relativity large tan$\beta$ will be accessible.  The
detectors with large R$_{coh}$ (e.g. Pb) are generally considerably
more sensitive than those with large R$_{inc}$ (e.g. CaF$_2$).
\vskip 0.6cm
\noindent
{\bf 8.  Sensitivity To Bounds on $\Omega_{\tilde{Z}_1}h^2$}
\vskip 0.4cm
In the preceeding discussion, we have assumed the bounds of Eq. (4) for
$\Omega_{\tilde{Z}_1}h^2$.  The results have some sensitivity to the
choice of endpoints and we examine this here.

As discussed in Sec. 6, small $\Omega_{\tilde{Z}_1}h^2$ generally leads
to large event rates, and R generally rises rapidly as
$\Omega_{\tilde{Z}_1}h^2$ decreases near its lower bound.  The largest
R, however, occurs for small $m_{\tilde{g}}$ (see e.g. Fig. 7), and
hence by the scaling relations discussed at the end of Sec. 4, for
small $m_{\tilde{W}_{1}}$.  However, there also exist cuts, e.g.
$m_{\tilde{W}_{1}}>$ 45 GeV, required by LEP data which forbids
$m_{\tilde{W}_{1}}$ from getting too small.  Thus one has a sharply
rising function R hitting an experimental constraint on the parameter
space.  This is the origin of the sharp peaks in the maximum event rate
curves of Figs. 11 and 12, i.e. the maximum event rate gets quite large
or not depending on whether or not the parameter point passes the
experimental cut.

The sensitivity of this effect is seen in Fig. 13 (for Pb) and Fig. 14
(for CaF$_2$) for the maximum event rates when one increases the
minimum value of $\Omega_{\tilde{Z}_1}h^2$
\vskip180pt
\noindent
Fig. 13  Maximum event rate as a function of $A_{t}/m_{0}$ for $\mu<$0
for Pb detector.  Solid curve is for $\Omega_{\tilde{Z}_1}h^2 >$0.10
and dot-dash for $\Omega_{\tilde{Z}_1}h^2 >$0.15.
\vskip 0.1cm
\noindent
from 0.10 to 0.15.  The sharp peaks get clipped off when the lower
bound on ($\Omega_{\tilde{Z}_1}h^2$) is increased.  There still
remains, however, a sizable portion of the parameter space
\vfill\eject
where R is large.
\vskip180pt
\noindent
Fig. 14  Maximum event rate as a function of $A_t/m_0$ for $\mu <$0 for
CaF$_2$ detector.  Dashed curve is for $\Omega_{\tilde{Z}_1}h^2 >$0.10
and solid curve for $\Omega_{\tilde{Z}_1}h^2 >$0.15.
\vskip 0.1cm
The minimum event rates are sensitive to the upper bound chosen for
$\Omega_{\tilde{Z}_1}h^2$.  This can be seen in Fig. 15.  The minimum
event rates increases by more than a factor of 10 as
($\Omega_{\tilde{Z}_1}h^2$)$_{max}$ is reduced from 0.35 to 0.20.  One
notes that the inflationary scenario favors a small Hubble constant,
i.e. h=0.5, so that the pre-
\vskip180pt
\noindent
Fig. 15  Minimum event rates as a function of
($\Omega_{\tilde{Z}_1}h^2$)$_{max}$ for Pb detector (solid curve) and
CaF$_2$ detector (dashed curve), $\mu <$0.
\vskip 0.1cm
\noindent
dicted age of the universe not be inconsistent with estimated ages of
globular star clusters.  Even if one assumed $\Omega_{\tilde{Z}_1} =1$
(i.e. no hot dark matter and negligible baryonic dark matter) this
would imply $\Omega_{\tilde{Z}_1}h^{2}$=0.25.  Thus the larger minimum
event rates of Fig. 15 may possibly be the correct choice.

The reason the minimum event rate increases with decreasing
($\Omega_{\tilde{Z}_1}h^2$)$_{max}$ is that the minimum rates occur for
the maximum values of $m_{\tilde{g}}$.  (As seen in Fig.  7 the event
rate drops with increasing $m_{\tilde{g}}$.)  Further, as discussed in
Sec. 7, Fig. 10, $m_{\tilde{g}}$ possesses a maximum value because of
the upper bound on $\Omega_{\tilde{Z}_1}h^2$.  Fig. 16 shows the
dependence of of ($m_{\tilde{g}}$)$_{max}$ on the maximum value of
$\Omega_{\tilde{Z}_1}h^2$.  As
\vskip180pt
\noindent
Fig. 16  Maximum value of $m_{\tilde{g}}$ as a function of the upper
bound on $\Omega_{\tilde{Z}_1}h^2$ for $\mu<$0, ${A_t}/m_{0}$=0.5,
tan$\beta$=6.  Results are insensitive to the values of $A_t$ and
tan$\beta$.
\vskip 0.1cm
\noindent
($\Omega h^2$)$_{max}$ is reduced from 0.35 to 0.20,
($m_{\tilde{g}}$)$_{max}$ drops from 750 GeV to 400 GeV.  The
inflationary scenario thus favors smaller values of $m_{\tilde{g}}$.
Such low $m_{\tilde{g}}$ implies that the gluino could be detected at
suggested energy upgrades of the Tevatron (e.g. at the
DiTevatron$^{12}$).
\vskip 0.6cm
\noindent
{\bf 9.  The b$\rightarrow$ s + $\gamma$ Decay}
\vskip 0.4cm
Recently, the CLEO Collaboration have measured the branching ratio for
the inclusive decay B$\rightarrow X_{s} + \gamma$:

$$BR\left( B\rightarrow X_{s} + \gamma\right)
  = \left (2.32\pm 0.51 \pm 0.29 \pm 0.32 \right )\times 10^{-4}\eqno(19)$$

\noindent
where the first error is statistical and the last two errors
systematic.  Combining all errors in Gaussian quadrature, one has in
the spectator approximation that BR(b$\rightarrow$s + $\gamma$) $\cong$
(2.32$\pm$0.67)$\times 10^{-4}$.  The b$\rightarrow$s + $\gamma$ decay
is of particular interest in that it begins at the loop level as it is
a FCNC process.  This means that SM and new physics effects enter at
the same loop level and one could expect large [i.e. O(1)] new physics
corrections to the SM predictions.  Thus the b$\rightarrow$s + $\gamma$
decay is an excellent process for detecting new physics.  We will
investigate what effects the current experimental value has on the SUSY
parameter space and hence on the expected dark matter detection rates.

The elementary diagrams at the electroweak scale $\mu = M_{W}$ are
shown in Fig. 17.  The $W^{-}-t$ intermediate state is the Standard
Model contribution, while the
\vskip150pt
\noindent
Fig. 17  Elementary penguin diagrams for b$\rightarrow$s + $\gamma$
decay at the electroweak scale $\mu = M_{W}$.  Only the third
generation quarks and squarks make a significant contribution.
\vskip 0.1cm
\noindent
$H^{-}-t$ and $\tilde W^{-}-\tilde{t}$ represent additional SUSY
contributions.  The interactions can be represented by an effective
Lagrangian for a transition magnetic dipole interaction$^{13}$:

$$L_{eff} = G_{F}V_{tb}V_{ts}^{*}m_{b}A_{\gamma}
\bar{s}_{L}\sigma^{\mu\nu} b_{R}F_{\mu\nu},\eqno(20)$$

\noindent
where the coefficient $A_{\gamma}$ can be evaluated in terms of the
basic parameters $m_{0}$, $m_{\tilde{g}}$, $A_{t}$, tan$\beta$.  In
order to calculate the decay rate, however, one must use the RG
equations to evaluate the amplitude at the b scale $\mu\approx m_{b}$.
This causes operator mixing with the gluonic transition magnetic moment
operator $\bar {s}_{L}\sigma^{\mu\nu}T^{a}b_{R}G_{\mu\nu}^{a}$ (where
$G_{\mu\nu}^{a}$, a=1...8 is the gluon field strength) and six four
quark operators.  It is convenient to consider the ratio

$$R={{BR(b\rightarrow s +\gamma)}\over{BR(b\rightarrow c+e+\bar\nu_{e})}}
\cong {{BR({B\rightarrow X_{s} +\gamma)}
    \over BR(B\rightarrow X_{c}+e+\bar{\nu}_{e})}}\eqno(21)$$

\noindent
since poorly known CKM matrix elements and $m_b$ factors cancel out in
the ratio.  One can recover the $b\rightarrow s+ \gamma$ rate then from
the experimental number of the charm semi-leptonic rate:
BR(B$\rightarrow X_{c} +e+\bar{\nu}_e$)=(10.7$\pm$0.5)\%.  To leading
order (LO) QCD the value of R is$^{13,14}$

$$R={{6\alpha}\over{\pi}}{{{|V_{ts}^{*}V_{tb}|^{2}}}\over{|V_{cb}|^{2}}}
{{{\left[{\eta^{16\over23}C_{7}(M_{W})-{8\over3}(\eta^{14\over23}
-\eta^{16\over23})C_{8}(M_{W})+C_{2}(M_{W})}\right]^{2}}}
\over{{I(z)\left[1-{2\over 3{\pi}}{{\alpha_{3}(M_{Z}})\over{\eta}}f(z)\right]}}
}\eqno(22)$$

\noindent
where $\eta = \alpha_{3} (M_{Z})/\alpha_{3}(m_{b})$=0.548,
z=$m_{c}/m_{b}$=0.316$\pm$0.013, I(z) is a phase space factor for the
b$\rightarrow$ce$\bar{\nu}_{e}$ decay, $C_{7}(C_{8})$ are the Wilson
coefficients for the photonic (gluonic) magnetic penguin operators and
$C_2$ comes from the operator mixing with the 4-quark operators.

There are a number of theoretical uncertainties in the above
calculation which can be sumarized as follows: (i) Errors in input
parameters i.e. $\alpha_{3}(M_Z)$, $m_b/m_c$, CKM factor, BR(B
$\rightarrow X_{c} e {\bar\nu}_{e}$); (ii) Errors in the spectator
approximation; (iii) There are large NLO (next to leading order) QCD
corrections.  This can be estimated by letting $\mu$ vary from
$m_{b}/2$ to 2 $m_{b}$ and are seen to be about $\pm$25\%; (iv) Heavy
mass threshold corrections in running the RGE from the t quark/squark,
$H^{-}$, $\tilde{W}$ threshold$^{15,16}$ down to $m_{b}$  These are
about 15\% for the t-quark and estimated to be $\pm$15\% for the SUSY
thresholds$^{16}$.  Thus current theory has an overall error of about
$\pm$30\%.

The current CLEO measurement of the b$\rightarrow s +\gamma$ rate has a
significant effect on the expected dark matter detector counting
rates.  Fig. 18 shows the expected b$\rightarrow s +\gamma$ rate for a
characterisitic choice of parameters$^{17}$.  The BR(b$\rightarrow s +\gamma$)
is increased when $\mu$ and $A_t$ have the same sign relative
to the value when $\mu$ and $A_t$ have the opposite sign.  (This effect
comes from the $\tilde{W}-\tilde{t}$ diagram of Fig. 17).  Thus regions
where $\mu$ and $A_t$ have the same sign can exceed the CLEO measured
value and such regions of parameter space are then experimentally
eliminated.  One sees from Fig. 18 also that the BR(b$\rightarrow s
+\gamma$) is largest when $m_0$ and $m_{\tilde{g}}$ are small, which we
also saw is the region when the dark matter counting rate R is
largest.  Thus one expects that the maximum values of R get eliminated
when $\mu$ and $A_t$
\vfill\eject
\noindent
have the same sign.  This may be seen in Fig. 19$^{18}$ (for $\mu<$ 0)
and Fig. 20$^{18}$ (for

\vskip210pt
\noindent
Fig. 18  BR(b$\rightarrow s +\gamma$) as a function of
$m_{\tilde{W}_1}$ for tan$\beta$=5.0, $|A_t$$/m_0|$=0.5, $m_t$=165 GeV,
$\alpha_{G}^{-1}$=24.11. Graphs (a) and (b) are for $A_t<0$,(c) and (d)
for $A_t>0$, while (a) and (c) are for $\mu >0$ and (b) and (d) for
$\mu<0$.
\vskip 0.1cm
\noindent
$\mu>$ 0).  There we have plotted R$_{Max}$ for the Pb detector without
the b$\rightarrow s +\gamma$ condition and R$_{Max}$ with parameter
points excluded when the predicted b$\rightarrow s +\gamma$
\vskip160pt
\noindent
Fig. 19  R$_{Max}$ vs $A_{t}/m_{0}$ for Pb detectors for $\mu<$0.  The
solid line is the expected rate without b$\rightarrow s +\gamma$
constraint and the dashed line is the rate with parameter points
excluded where the predicted b$\rightarrow s +\gamma$ rate lies outside
95\% C.L. bound of the experimental value of Eq. (19).
\vskip 0.1cm
\noindent
 rate exceeds the 95\% C.L. of Eq. (19).  We see that when $\mu$ and
 $A_t$ have the same sign, the maximum event rate drops sharply, well
below what could be observable

\vskip150pt
\noindent
Fig. 20  Same as Fig. 19 for $\mu>$0.
\vskip 0.1cm
\noindent
in the forseeable future.  However, the COBE bounds of Eq. (4) turn out
to allow mostly $A_{t}>$0.  Thus the major effect of the CLEO
measurement of b$\rightarrow s +\gamma$ is for $\mu>$0 where for most
of the parameter space, the event rate will be very small and hence
unobservable.  However, for $\mu<$0, the b$\rightarrow s +\gamma$
measurement does not effect the expected rates very much, and large
dark matter event rates are still possible.
\vskip 0.6cm
\noindent
{\bf 10.  Proton Decay}
\vskip 0.4cm
The preceeding discussion has been for a generic supergravity Gut model
described in Sec. 3 for the parameter domain of Eq. (18).  Results are
generally independent of the Gut group and Gut physics provided that
Gut threshold effects are not so strong that they prevent grand
unification from occuring at $M_{G}\simeq10^{16}$ GeV.

Proton decay is characteristic of all supergravity Gut models except
for the flipped SU(5) model$^{19}$.  Further supersymmetry generally
implies a unique dominant decay mode:

$$p\rightarrow \bar\nu + K^{+}\eqno(23)$$

\noindent
Thus the observation of this decay would not only indicate the validity
of grand unification but also of supersymmetry.  One can suppress this
decay by specially tailoring the form of the Gut Higgs sector, but this
generally requires some awkward fine tuning.  We consider here
``SU(5)-type'' proton decay$^{20}$ which arises via the exchange of a
superheavy color triplet Higgsino $\tilde{H}_{3}$, Fig. 21.  (This can
happen in SU(5), O(10), $E_6$ etc. Gut groups.) The decay rate can be
written as,

$$\Gamma \left ( p\rightarrow \bar\nu + K^{+}\right )
  = {{const} \over {M_{H_{3}}^{2}}}|B|^2,\eqno(24)$$
\noindent
where B is the loop amplitude.  The current experimental bound$^{21}$,

\vskip150pt
\noindent
Fig. 21  Example of proton decay diagram for $p\rightarrow \bar{\nu} +
K^{+}$ for supersymmetric grand unification.  The $\tilde{H}_{3}$
vertices violate baryon and lepton number.
\vskip 0.1cm
\noindent
$\tau\left (p\rightarrow \bar{\nu} + K \right ) >$1.0$\times 10^{32}$ y,
then leads to a bound on $B$

$$B \buildrel{<}\over{\sim}100 \left (M_{H_{3}}/M_{G}\right ) GeV ^{-1};
{}~~ M_{G}=2\times 10^{32} GeV\eqno(25)$$

\noindent
We restrict $M_{H_{3}}$ to obey $M_{H_{3}}/M_{G}<10$ in the following
so that the Gut scale be disjoint from the Planck scale.  (For larger
$M_{H_{3}}$ one might expect large Planck physics corrections to enter,
the nature of which are not known.)  B can be expressed in terms of
$m_{\tilde{W}}$, $m_{\tilde{q}}$ etc. and hence by the RG equations in
terms of the four basic parameters $m_0$, $m_{\tilde{q}}$, $A_t$,
tan$\beta$ and the sign of $\mu$.  Thus the condition on B is a
constraint on the parameter space.

The second generation dominates the loop of Fig. 21 and to a rough
approximation on has

$$B_{2}\approx -{{2\alpha_{2}}\over{\alpha_{3}sin2\beta}}
{{m_{\tilde{g}}}\over{m_{\tilde{q}}^{2}}}\times {10^{6}} GeV^{-1}\eqno(26)$$

\noindent
where $m_{\tilde{q}}^{2}\cong m_{0}^{2}+0.6 m_{\tilde{g}}^{2}$.  Thus
the proton decay bounds of Eq. (25) imply small $m_{\tilde{g}}$, large
$m_0$ e.g. $m_{0}>m_{\tilde{g}}$, and small tan$\beta$ i.e.
tan$\beta\leq$10.  One may satisfy both the COBE constraints on
$\Omega_{\tilde{Z}_{1}}h^{2}$ of Eq. (4) simultaneously with the proton
decay constraints of Eq. (25) even though $m_0$ must be large (which
usually is a region of small relic $\tilde{Z}_{1}$ annihilation).  This
is possible since, as discussed in Sec. 5, when $m_{\tilde{g}}\leq$450 GeV,
(the region also required by Eq. (25) for proton decay) large
relic $\tilde{Z}_{1}$ annihilation can occur due to the presence of h
or Z poles.  In fact, one finds that in the vacinity of these poles,
when $m_0$ is small (e.g. $m_{0} < m_{\tilde{g}}$) too much
annihilation occurs [i.e. $\Omega_{\tilde{Z}_{1}}h^{2}$=O(10$^{-2}$)]
and one must increase $m_0$ to satisfy the lower bound of Eq. (4).
These are then the domains that also satisfy the proton decay
constraint.  Thus in order to find the parameter space region which
simultaneously satisfies the dark matter and proton decay constraints,
it is essential to treat the calculation of $\tilde{Z}_{1}$ relic
density in the accurate fashion discussed in Sec. 5.  One finds then
that the parameter points satisfying the simultaneous constraints
require

$$m_{\tilde{g}}\leq 375 GeV;~~ m_{0}\geq 500 GeV; ~~tan\beta \leq 10$$
$$0.0\leq A_{t}/m_{0}\leq 0.5\eqno(27)$$
\noindent
for a t quark of mass 165 GeV.

One may next ask whether the CLEO measurement of the b$\rightarrow s +
\gamma$ decay effects this result.  One finds, however, that 95\% of
the parameter points which simultaneously satisfy the dark matter and
proton decay constraints, predict a b$\rightarrow s + \gamma$ branching
ratio in the LO that is within the 90\% C.L. of the experimental value
of Eq. (19).  Thus, at the current experimental and theoretical
accuracy, the b$\rightarrow s + \gamma$ decay does not effect the
proton decay predictions.

However, the proton decay constraint does effect the expected dark
matter event rates, and the maximum event rates are significantly
reduced since tan$\beta$ is small and $m_0$ is large.  Parameter points
which simultaneously satisfy COBE, the b$\rightarrow s + \gamma$ and
proton decay constraints lead to event rates of size
R=O(10$^{-3}$-10$^{-4}$) events/kg da.  Thus if the next round of
proton decay experiments (Super Kamiokande, ICARUS) were to actually
detect proton decay, the present theory implies that $\tilde{Z}_{1}$
dark matter is beyond the ability of current detectors to discover by
direct detection.
\vskip 0.6cm
\noindent
{\bf 11.  Conclusions}
\vskip 0.4cm
Studying non high energy accelerator phenomena such as dark matter,
proton decay, the b$\rightarrow s + \gamma$ decay is useful in limiting
the SUSY parameter space.  We have examined within the framework of the
minimal supergravity model (MSGM) the ability of direct detection of
dark matter when the relic density obeys Eq. (4).
\vskip 0.1cm

\item{(i)}  The detectors which are most sensitive to coherrent
$\tilde{Z}_{1}$ scattering (e.g. the Pb detector) are better than the
detectors most sensitive to incoherrent (spin-dependent) scattering,
and the heavier the nucleus, the more sensitive the detector.

\item{(ii)}  With a future sensitivity of R $>$ 0.01 events/kg da, a
reasonable amount, though not all, of the parameter space will be
accessible to dark matter detectors.

\item{(iii)}  Raising the lower bound on $\Omega_{\tilde{Z}_{1}}h^2$
decreases the maximum event rate (Fig. 13, 14), and lowering the upper
bound increases the minimum event rate (Fig. 15).  The upper bound,
$\Omega_{\tilde{Z}_{1}}h^{2}$=0.35, also determines an upper bound on
$m_{\tilde{g}}$ of 750 GeV.  Thus if this upper bound is lowered (as
suggested by the inflationary scenario without a cosmological constant)
the gluino would become more accessible to accelerator detection.

\item{(iv)}  The largest event rates occur at large tan$\beta$ and
small $m_{\tilde{g}}$ and small $m_0$ (Figs. 7-9).  Thus detectors are
most sensitive to these domains.  Models with very large tan$\beta$
(i.e. tan$\beta\approx$50) may therefore be testable by planned
detectors.

\item{(v)}  The predicted b$\rightarrow s + \gamma$ decay is large in
the same region of parameter space (small $m_{\tilde{g}}$, small $m_0$)
where dark matter event R is large when $\mu$ and $A_t$ have the same
sign.  The current experimental rate for this decay thus eliminates
part of the parameter space where R is large.  Since the relic density
constraint eliminates most of the $A_{t}<0$ part of the parameter
space, the b$\rightarrow s + \gamma$ constraint significantly reduces
the expected event rate for $\mu > 0$, but does not effect the $\mu <
0$ event rates a great deal, and large event rates can still occur for
$\mu < 0$.

\item{(vi)}  For models possessing in addition SU(5)-type proton decay,
there remain regions in parameter space, Eq. (27), satisfying both the
relic density constraint and current proton decay bounds.  These points
also are within the 90\% C.L. bounds of the current b$\rightarrow s +
\gamma$ decay rate.  However, since proton decay favors large $m_0$ and
tan$\beta \leq$10, the predicted dark matter event rates are all for R
$<$0.01 event/kg da.  Hence models with SU(5)-type proton decay predict
that $\tilde{Z}_{1}$ dark matter will be inaccessible to current
detectors.
\vskip 0.6cm
\noindent
{\bf Acknowledgements}
\vskip 0.4cm
This work was supported in part by NSF grant numbers PHY-9411543 and
PHY-19306906.

\vskip 0.6cm
\noindent
{\bf References}
\vskip 0.4cm
\item{1.}
P.F. Smith and J.D. Lewin, Phys. Rep. $\underline{187}$, 203 (1990).
\item{2.}
R. Sancisi and T.S. van Albada, ``Dark Matter in the Universe'', eds.
J. Kormendy and G. Knapp (Reidel, Dordrect, 1987).
\item{3.}
M. Kamionkowski, Phys. Rev. $\underline{D44}$, 3021 (1991); F. Halzen,
M. Kamionkowski and T. Steltzer, Phys. Rev. $\underline{D45}$, 4439
(1992); R. Gandhi, J.L. Lopez, D.V. Nanopoulos, K. Yuan and A.
Zichichi, Phys. Rev. $\underline{D49}$, 3691 (1994); A. Bottino, N.
Fornengo, G. Mignola and L.  Moscoso, DFTT 34/94 (to appear
Astroparticle Physics).
\item{4.}
A.H. Chamseddine, R. Arnowitt and P. Nath, Phys. Rev. Lett.
$\underline{49}$, 970 (1982).  For reviews see P. Nath, R. Arnowitt and
A.H. Chamsedine, ``Applied N=1 Supergravity'' (World Scientific,
Singapore 1984); H.P. Nilles, Phys. Rep. $\underline{110}$, 1 (1984);
R. Arnowitt and P. Nath, Proc. of VII J.A. Swieca Summer School, Brazil
(World Scientific, Singapore, 1994).
\item{5.}
R. Arnowitt and P. Nath, Phys. Rev. Lett. $\underline{69}$, 725 (1992);
P. Nath and R. Arnowitt, Phys. Lett. $\underline{B289}$, 368 (1992).
\item{6.}
B.W. Lee and S. Weinberg, Phys. Rev. Lett. $\underline{39}$, 165
(1977); D.A. Dicus, E. Kolb and V. Teplitz, Phys. Rev. Lett.
$\underline{50}$, 1419 (1983), J. Ellis, J.S. Hagelin, D.V.
Nanopoulos, K. Olive and M. Srednicki, Nucl. Phys. $\underline{B238}$,
453 (1984).
\item{7.}
K. Griest and D. Seckel, Phys. Rev. $\underline {D43}$, 3191 (1991); P.
Gondolo and G. Gelmini, Nucl. Phys. $\underline{360}$, 145 (1991).
\item{8.}
P. Nath and R. Arnowitt, Phys. Rev. Lett. $\underline{70}$, 3696 (1993).
\item{9.}
W. Goodman and E. Witten, Phys. Rev. $\underline{D31}$, 3059 (1985); K.
Greist, Phys. Rev.  Lett. $\underline{62}$, 666 (1988); Phys. Rev.
$\underline{D38}$, 2357 (1988); R. Barbieri, M. Frigini and G.F.
Giudice, Nucl. Phys. $\underline{B313}$, 725 (1989); M. Srednicki, and
R. Watkins, Phys. Rev. Lett. $\underline{B225}$, 140 (1989); G.F.
Giudice and E. Roulet, Nucl.  Phys. $\underline{B316}$, 429 (1989); J.
Ellis and R. Flores, Phys. Lett. $\underline{B263}$, 259 (1991);
$\underline{B300}$, 175 (1993); Nucl. Phys. $\underline{B400}$, 25 (1993).
\item{10.}
M. Kamionkowski, Phys. Rev. $\underline{D44}$, 3021 (1991); M. Drees
and M. Nojiri, Phys.  Rev. $\underline{D48}$, 3483 (1993).
\item{11.}
R. Arnowitt and P. Nath, CERN-TH.  7362/94-CTP-TAMU-37/94-NUB-TH-3098/94;
P. Nath and R.  Arnowitt, CERN-TH. 7363/94-NUB-TH-3099/94-CTP-TAMU-38/94..
\item{12.}
T. Kamon, J.L. Lopez, P. McIntyre and J. White, Phys.Rev. $\underline{D50}$,
5676 (1994).
\item{13.}
S. Bertolini, F. Borzumati and A. Masiero, Nucl. Phys.
$\underline{B294}$, 321 (1987); B. Grinstein, R. Springer and M. Wise,
Nucl. Phys. $\underline{B339}$, 269 (1990); S. Bertolini, F.
Borzumati, A. Masiero and G. Ridolfi, Nucl. Phys. $\underline{B353}$,
591 (1991).
\item{14.}
M. Misiak, Phys. Lett. $\underline{B269}$, 161 (1991); Nucl. Phys.
$\underline{B393}$, 23 (1993); M. Ciuchini, E. Franco, G. Martinelli,
L. Reina and L. Silvestrini, Phys. Lett. $\underline{B316}$, 127
(1993); A.J. Buras, M. Misiak, M. M\"unz and S. Pokorski, MPI-Ph/93-77
(1993); M. Misiak, Phys. Lett. $\underline{321B}$, 113 (1994).
\item{15.}
P. Cho and B. Grinstein, Nucl. Phys. $\underline{B365}$, 279 (1991); E
(in print);C.S.  Gao, J.L. Hu, C.D. Lu, and Z.M. Qiu, Beijing preprint
(1993).
\item{16.}
H. Anlauf, SLAC-PUB-6525 (1994) (hepph-9406286).
\item{17.}
J. Wu, R. Arnowitt and P. Nath, CERN-TH.7316-CTP-TAMU-03/94-NUB-TH.3092/94.
\item{18.}
R. Arnowitt and P. Nath, in preparation.
\item{19.}
I. Antoniadis, J. Ellis, J. Hagelin and D.V. Nanopoulos, Phys. Lett.
$\underline{B208}$, 209 (1988).
\item{20.}
J. Ellis, D.V. Nanopoulos and S. Rudaz, Nucl. Phys. $\underline{B202}$,
43 (1982); R. Arnowitt, A.H. Chamseddine and P. Nath, Phys. Lett.
$\underline{B156}$, 215 (1985); P. Nath, R. Arnowitt, and A.H.
Chamseddine, Phys. Rev. $\underline{D32}$, 2348 (1985); S. Kelley, J.
Lopez, D.V.  Nanopoulos and K. Yuan, Phys. Lett. $\underline{B272}$,
423 (1991); R. Arnowitt and P. Nath, Phys. Rev. Lett. $\underline{69}$,
725 (1992); P. Nath and R. Arnowitt, Phys. Lett. $\underline{B289}$,
368 (1992); R. Arnowitt and P. Nath, Phys. Rev. $\underline{D49}$, 1479
(1994).
\item{21.}
Particle Data Group, Phys. Rev. $\underline{D50}$, Part 1 (1994).

\end{document}